\documentclass[12pt]{article}
\usepackage[dvips]{graphicx}
\usepackage{amssymb,psfrag}
\title{Random Wandering Around
Homoclinic-like Manifolds in Symplectic Map Chain}
\author{Shin-itiro Goto\footnote{e-mail:~sgoto@allegro.phys.nagoya-u.ac.jp}
\quad , Kazuhiro Nozaki\quad  and Hiroyasu Yamada  \\
{\it Department of Physics,Nagoya University,Nagoya 464-8602,Japan}}
\date{}
\makeindex

\begin{document}
\maketitle

\newcommand{\beq}{\begin{equation}}
\newcommand{\beqa}{\begin{eqnarray}}
\newcommand{\eeq}{\end{equation}}
\newcommand{\eeqa}{\end{eqnarray}}
\newcommand{\non}{\nonumber}
\newcommand{\lb}{\label}
\newcommand{\fr}[1]{(\ref{#1})}
\newcommand{\cc}{\mbox{c.c.}}
\newcommand{\nr}{\mbox{n.r.}}
\newcommand{\tx}{\widetilde{x}}
\newcommand{\tg}{\widetilde{g}}
\newcommand{\hx}{\widehat{x}}
\newcommand{\tA}{\widetilde A}
\newcommand{\tB}{\widetilde B}
\newcommand{\tK}{\widetilde K}
\newcommand{\tc}{\widetilde c}
\newcommand{\tAc}{{\widetilde A}^{*}}
\newcommand{\tphi}{{\widetilde \phi}}
\newcommand{\btA}{\mbox{\boldmath {$\widetilde A$}}}
\newcommand{\bA}{\mbox{\boldmath {$A$}}}
\newcommand{\bC}{\mbox{\boldmath {$C$}}}
\newcommand{\bu}{\mbox{\boldmath {$u$}}}
\newcommand{\bN}{\mbox{\boldmath {$N$}}}
\newcommand{\bZ}{\mbox{\boldmath {$Z$}}}
\newcommand{\bR}{\mbox{\boldmath {$R$}}}
\newcommand{\ve}{{\varepsilon}}
\newcommand{\sabun}{\bigtriangleup_j^2}
\newcommand{\aCos}{\mbox{Cos}^{-1}}
\newcommand{\aTan}{\mbox{Tan}^{-1}}
\newcommand{\sech}{\mbox{sech}}
\begin{abstract}
We present a method to construct a symplecticity preserving
   renormalization group
map of a chain of weakly nonlinear symplectic maps
and  obtain a general reduced symplectic map describing its
   long-time behaviour. It is found that the modulational instability in
   the reduced map triggers random wandering of orbits around
   some homoclinic-like manifolds, which is understood as
the Bernoulli shifts.

\end{abstract}

\pagebreak

\section{Introduction}
\qquad
Hamiltonian systems with more than one degree of freedom
can exhibit trajectories with complex behavior.
Discrete symplectic maps are of interest
since  the Poincar\'e surface of section leads naturally to a mapping
of the dynamical trajectory onto a subspace of the phase space of
the Hamiltonian flow. Symplectic maps allow much easier numerical
calculations of the motion than Hamiltonian flows, which are particularly
convenient for studying systems of many degrees of freedom.
In a two-dimensional symplectic mapping, which corresponds to
Poincar\'e mapping of continuous-time Hamiltonian systems with two
degrees of freedom, the main ingredients of
this chaotic motion are hyperbolic type orbits and their invariant
manifolds. The existence of transversal intersection between stable and
unstable manifolds leads to non integrability of Hamiltonian systems.
The area of the lobe enclosed by stable and unstable manifolds
represents the flux from inside ({\it resp}. outside)
the separatrix to outside ({\it resp}. inside).
The manifolds which have codimension one
separate the space into disjoint regions.

For symplectic mappings of four or more dimensions,
there are some studies about the homoclinic bifurcation near  fully
hyperbolic fixed points \cite{GS95} \cite{HNK99-101}\cite{HNK99-102}.
Although the dimension of unstable and stable manifolds of a fully
hyperbolic fixed point takes  a maximum possible value, the
transversal intersection occurs on a nearly one-dimensional homoclinic
submanifold and exponentially small splitting of the homoclinic manifold
leads to very weak chaos, which is almost invisible through naive
numerical experiments.

   A chain of  weakly coupled nonlinear oscillators may present a model for
   dynamics near a fully elliptic fixed point.
The equilibrium state of almost independent oscillators corresponds to an
fully elliptic fixed point in the phase space and periodic orbits or tori
encircle
the elliptic fixed point.  Resonant tori  may be disrupted
by the well-known resonance mechanism.
If a torus is hyperbolic,  unstable and stable manifolds attach to the
torus as a ``whisker''  \cite{Arn64}.
   Some dynamics of  such whiskers  attached to tori  is studied near
the resonance junction  by means of the naive averaging method
   \cite{GN99} .

In the case of a chain of weakly coupled nonlinear oscillators, where both coupling and
nonlinearity are weak,  tori  may be destabilized by another mechanism
called the modulational instability.
Here, we
study about such modulationally unstable tori and their unstable manifolds.
We start with a chain of symplectic maps, each of which represents
a weakly nonlinear oscillator. In order to avoid special dependence
of results on the specific form of  nonlinearity,
our investigation is focused on long-time asymptotic behavior of
   dynamics of the system. However, the conventional asymptotic
methods such as the averaging method or the method of the multiple-time
scales may not be immediately applicable to discrete systems.
Instead, we employ the perturbative renormalization group (RG) method
   \cite{CGO96}\cite{GMN99},
which can apply to a discrete system  and leads to a
reduced map as the RG map \cite{GN01Prog}.
However, a naive reduced map does not
preserve  symplectic symmetry and fails to
describe  long time behavior of the original symplectic map.
For two-dimensional maps  or special four-dimensional maps, the symplecticity
   is  recovered through a simple ``regularization'' procedure
   \cite{GN01JPSJ}.
For higher dimensional symplectic maps, we do not have a general procedure
of ``regularization''. The first purpose of this paper is to develop
a general procedure of ``regularization'' of the RG map for  a higher 
dimensional
symplectic map. Our regularization procedure is based on the symplectic
  integration method for a continuous Hamiltonian flow and includes the
  previous regularization procedure as a special case. By means of this
  regularization procedure, we can
   derive  a reduced symplectic map as the RG
map, which asymptotically approximates a chain of  weakly nonlinear symplectic
maps.¡¡

The reduced map obtained here is general in the similar sense that
the nonlinear Schr\"odinger equation asymptotically describes
a general weakly nonlinear dispersive wave.
In fact, the reduced map is found to be a space-time discrete version of
nonlinear Schr\"odinger equation.
Since the reduced map is not integrable, while its continuous limit is
integrable, it is worth studying chaotic solutions of the reduced map,
which is the second purpose of this paper. The reduced map has a periodic
solution, which is destabilized by modulational perturbation under
some conditions (the modulational instability).  Then, the periodic
solution becomes hyperbolic and is associated with unstable and stable
manifolds, whose transversal intersection is the origin of chaos
(the homoclinic chaos).
It should be mentioned that various discrete versions of the nonlinear
Schr\"odinger equation are studied in connection with numerically induced
chaos of the nonlinear Schr\"odinger equation \cite{AHS96}\cite{WS92} .
There, the systems of a large degrees of freedom are investigated
in connection with numerical integration of the nonlinear Schr\"odinger
equation and
only few attentions are directed to behavior of chaotic solutions themselves.

Here, we are mainly interesting in chaotic solutions near the onset of
symplectic
chaos, where the phase space of the map has a dimension higher than two.
Near the onset of the homoclinic chaos, an unstable manifold is found to
be very close to
a stable manifold and a homoclinic-like structure is approximately preserved.
In a two-dimensional symplectic map,
the homoclinic chaos close to the onset is hardly seen since  a splitting
distance between unstable and stable manifolds is small beyond all order of
the perturbation parameter. However, in a higher dimensional map considered
in this paper,
the weak chaos is shown to be visualized as a random rotation of
some homoclinic-like orbits, which is characterized by
the Bernoulli shift.

In the section 2,  some reduced maps are derived from a chain of symplectic
   maps as  the regularlized RG map.
   The conditions of the modullational instability in the reduced maps
   are presented  and unstable manifolds of unstable periodic orbits
   are depicted  in the cases of two, three and four
sites in the section 3.  In the section 4, a random rotation of
homoclinic-like orbits
is observed for three- and four- site maps, which is interpreted
   as the the Bernoulli shift of three or four numbers.

\section{Symplectic Map Chain and  Reduced Map}
In this section, we consider  a  symplectic map
chain  and present a method to construct a symplecticity preserving
   renormalization group map ( a reduced symplectic map) from the
   symplectic map chain.
\subsection{Symplecticity Preserving Renormalization Group method}
We give a method to derive a symplecticity preserving
reduced map from a nonlinear symplectic map of three or more sites,
which
is non-integrable even in the time-continuous limit.
\beqa
x_j^{n+1}&=&x_j^{n}+\tau p_j^{n+1},\lb{nl-model-x}\\
p_j^{n+1}&=&p_j^{n}+\tau \bigg(-\Omega^2 x_j^{n}
           +\ve\bigg\{\nu (x_{j+1}^n-2x_j^n+x_{j-1}^n)-\alpha (x_j^n)^3 \bigg\}
              \bigg),\lb{nl-model-p}
\eeqa
with $N(\geq3)$-periodic boundary condition
: $(x_{j+N}^n,p_{j+N}^n)=(x_{j}^n,p_{j}^n)$.
Here $x_j^n$ and $p_j^n$ are  real canonical variables
on site $j$ at time $n$, while
   $\Omega , \alpha, \nu$, and $\tau$ are
real parameters and
$\ve$ is a small perturbation parameter, $0<\ve\ll 1$.
Eliminating  $p$ -variables from Eqs. \fr{nl-model-x} and
\fr{nl-model-p},
we obtain a map for $x$-variables:
\beq
x_j^{n+1}-2\cos (\theta) x_j^n-x_j^{n-1}
=\ve\tau^2\bigg\{\nu (x_{j+1}^n-2x_j^n+x_{j-1}^n)-\alpha (x_j^n)^3\bigg\},
\lb{nl-model-x-o}
\eeq
where $\cos(\theta)=(1-\Omega^2\tau^2/2)$.
Let us construct a  reduced map for $0 <\ve\ll 1$ by means of
the RG method.
Expanding a solution of Eq. \fr{nl-model-x-o} as
$x_j^n=x_j^{n(0)} + \ve x_j^{n(1)}+{\cal O}(\ve^2), \quad j=1,2,.. $,
we have
\beqa
x_j^{n(0)}&=&A_j\exp(-i\theta n)+\cc,\non\\
{\cal L}x_j^{n(1)}&=&\tau^2
\bigg\{
\nu \sabun A_j-3\alpha |A_j|^2 A_j
\bigg\}\exp(-i\theta n)+\cc,\non
\eeqa
where
\beqa
{\cal L} x_j^{n(1)}&\equiv&
   x_j^{n+1(1)}-2\cos\theta\cdot x_j^{n(1)}+x_j^{n-1(1)},\lb{def-Lx}\\
\sabun A_j^n&\equiv&A_{j+1}^n-2 A_j^n+A_{j-1}^n,
\eeqa
and $\cc$ denotes complex conjugate to the preceding terms;
$A_j$ is an integration constant associated with a site $j$.
The first order solution has a secular term with respect to $n$.
$$
x_j^{n(1)}=n\frac{i\tau^2}{2\sin\theta}
\bigg\{
\nu \sabun A_j-3\alpha |A_j|^2  A_j
\bigg\}\exp(-i\theta n) +\cc+\nr,
$$
and $\nr$ denotes non-resonant terms. 
In order to remove this secular term, we define a renormalization
transformation $A_j\mapsto A_j^n$ by
\beq
A_j^n\equiv A_j+\ve\bigg\{n\frac{i\tau^2}{2\sin\theta}
\bigg(\nu\sabun A_j-3\alpha |A_j|^2 A_j
\bigg)\bigg\}+{\cal O}(\ve^2).
\lb{nl-model-RGT}
\eeq
A discrete version of the  RG equation is constructed
by taking  difference of $A_j^n$,
\beq
A_j^{n+1}-A_j^n=
\ve\bigg\{\frac{i\tau^2}{2\sin\theta}\bigg(\nu\sabun A_j-3\alpha |A_j|^2 A_j
\bigg)\bigg\}.
\lb{nl-model-diff-Ajn}
\eeq
Substituting the expression for $A_j$ in terms of $A_j^n$  defined by
Eq. \fr{nl-model-RGT} into Eq.
\fr{nl-model-diff-Ajn}, we can eliminate the secular term up to
${\cal O}(\ve)$ and obtain a naive RG map
\beq
A_j^{n+1}=A_j^n+
\ve\bigg\{\frac{i\tau^2}{2\sin\theta}\bigg(\nu\sabun A_j^n
-3\alpha |A_j^n|^2 A_j^n
\bigg)\bigg\},
\lb{nl-model-nRGM}
\eeq
or
\beq
\frac{A_j^{n+1}-A_j^n}{\tau}=
\ve\bigg\{\frac{i\tau}{2\sin\theta}\bigg(\nu\sabun A_j^n
-3\alpha |A_j^n|^2 A_j^n
\bigg)\bigg\}.
\lb{nl-model-nRGM-Euler}
\eeq
This naive RG map does not preserve symplectic symmetry.
To recover symplecticity of Eq. \fr{nl-model-nRGM},
we apply  the symplectic integration
method to the continuous-time limit ($\tau\to 0$) of
the naive RG map \fr{nl-model-nRGM-Euler}.
The continuous-time limit takes the following symplectic form,
which is also derived from the continuous-time limit of
Eqs. \fr{nl-model-x} and \fr{nl-model-p} in the
Appendix A,
\beq
\frac{dA_j}{dt}=i\ve\frac{1}{2\Omega}
\bigg(\nu\sabun A_j^n-3\alpha |A_j^n|^2 A_j^n
\bigg)=\frac{\partial H^{RG}}{\partial A_j^*} ,\qquad
\frac{dA_j^*}{dt}=-\frac{\partial H^{RG}}{\partial A_j}.
\lb{conti_naiveRG}
\eeq
Here, $\displaystyle{\Omega=\frac{\sin(\theta)}{\tau}}$ as $\tau\to 0$ and
$H^{RG}$ is given by
\beqa
H^{RG}&=&H^{\nu}+H^{\alpha},\non\\
H^{\nu}&=&-\frac{i\ve\nu}{2\Omega}\sum_j|A_{j+1}-A_j|^2,\non\\
H^{\alpha}&=&\frac{3i\ve\alpha}{4\Omega}\sum_j|A_j|^2.\non
\eeqa
The Hamiltonian flow defind by $H^{\alpha}$ can be solved  as
$$
A_j(t+\tau')=
A_j(t)\exp\bigg(-\frac{3i\ve\alpha}{2\Omega}|A_j(t)|^2\tau'\bigg), 
$$
where $\tau'$ is an arbitrary constant and we have
\beq
A_j^{n+1}=A_j^n\exp\bigg(-\frac{3i\ve\alpha}{2\Omega}|A_j^n|^2\tau\bigg).
\lb{Ha-flow}
\eeq
The flow defined by $H^{\nu}$ can  be discretized  by the symplectic implicit
midpoint rule \cite{HVA94}:  a general Hamiltonian flow
$$
\frac{dz}{dt}=f(z),
$$
is discretized as
$$
z^{n+1}=z^{n}+\tau f\bigg(\frac{1}{2}(z^{n+1}+z^{n})\bigg).
$$
In our case, the scheme gives
\beq
A_j^{n+1}=A_j^{n}+\tau\frac{i\ve\nu}{2\Omega}\frac{1}{2}
\bigg(\sabun A_j^{n+1}+\sabun A_j^{n}\bigg),
\non
\eeq
or a temporally explicit form
\beq
\bigg(1-i\tau\frac{i\ve\nu}{2\Omega}\frac{1}{2}\sabun\bigg)A_j^{n+1}
=
\bigg(1+i\tau\frac{i\ve\nu}{2\Omega}\frac{1}{2}\sabun\bigg)A_j^{n}.
\lb{Hn-flow}
\eeq

Combining  two symplectic transformations \fr{Ha-flow} and \fr{Hn-flow} ,
we get a symplectic scheme for $H=H^{\nu}+H^{\alpha}$ as
$$
\bigg(1-i\tau\frac{i\ve\nu}{2\Omega}\frac{1}{2}\sabun\bigg)A_j^{n+1}
=
\bigg(1+i\tau\frac{i\ve\nu}{2\Omega}\frac{1}{2}\sabun\bigg)
\exp\bigg(-\frac{3i\ve\alpha}{2\Omega}|A_j^n|^2\tau\bigg)A_j^n.
$$
For brevity, we rewrite the above symplectic RG map as
\beq
\bigg(1-iT\sabun\bigg)A_j^{n+1}
=\bigg(1+iT\sabun\bigg)
\exp\bigg(iQ|A_j^n|^2\bigg)A_j^n,
\lb{nl-model-RRGM}
\eeq
where
$$
T\equiv\ve\frac{\nu\tau^2}{4\sin\theta}\in\bR,\qquad
Q\equiv \ve\frac{-3\alpha\tau^2}{2\sin\theta}\in\bR.
$$
Which has a conserved quantity $\sum_j |A_j^n|^2$ (see Ref. \cite{WH86}).
Both temporally and spatially explicit form of Eq.\fr{nl-model-RRGM}
is possible. For instance, the three-cite map is written as
$$
\left(
\begin{array}{c}
A_1^{n+1} \\
A_2^{n+1} \\
A_3^{n+1}
\end{array}
\right)
=\frac{1-3iT}{1+9T^2}
\left(
\begin{array}{ccc}
1-iT & 2iT & 2iT \\
2iT  &1-iT &2iT \\
2iT  &2iT  &1-iT
\end{array}
\right)
\left(
\begin{array}{c}
B_1^{n} \\
B_2^{n} \\
B_3^{n}
\end{array}
\right),
$$
where
$$
B_j^n\equiv\exp(iQ|A_j^n|^2)A_j^n.
$$
An approximate solution of the original map \fr{nl-model-x} and \fr{nl-model-p}
  is expressed
in terms of the renormalized amplitude  as $x_j^n\approx A_j^n\exp (-i\theta n)
+\cc$
If the coupling parameter $\nu$ vanishes, the present regularization process is
nothing but a simple exponentiation procedure introduced in  \cite{GN01JPSJ}.
Note that the Eq.\fr{nl-model-RRGM} with $N\gg 1$  is identical to
a difference scheme for the nonlinear Schr\"odinger equation \cite{WH86}.
In fact, the time-continuous limit of the RG map \fr{nl-model-RRGM}
   is just a spatially discretized nonlinear Schr\"odinger equation (see
Appendix A).
The map \fr{nl-model-RRGM} also admits some
oscillating solutions including a spatially uniform one.

\subsection{Nonlinear symplectic map with two sites}
Let us consider a nonlinear symplectic map with two sites.
This system is separately discussed  since
the continuous-time limit ($\tau\to 0$) of its RG map
is integrable.
\beqa
x_1^{n+1}&=&x_1^{n}+\tau p_1^{n+1},\qquad
x_2^{n+1}=x_2^{n}+\tau p_2^{n+1},\lb{2disc-orig-model-1}\\
p_1^{n+1}&=&p_1^{n}+\tau \bigg\{-\Omega^2 x_1^{n}
     +\ve\bigg(\nu (x_2 -x_1) -\alpha
x_1^3\bigg)\bigg\},\lb{2disc-orig-model-2}\\
p_2^{n+1}&=&p_2^{n}+\tau \bigg\{-\Omega^2 x_2^{n}
    +\ve\bigg(\nu (x_1 -x_2) -\alpha x_2^3\bigg)\bigg\}.\lb{2disc-orig-model-3}
\eeqa
We rewrite Eqs. \fr{2disc-orig-model-1} -- \fr{2disc-orig-model-3}
   as follows,
\beqa
x_1^{n+1}-2\cos\theta\cdot x_1^n+x_1^{n-1}&=&\ve\tau^2
\bigg\{\nu (x_2^n-x_1^n)-\alpha (x_1^n)^3 \bigg\},\non\\
x_2^{n+1}-2\cos\theta\cdot x_2^n+x_2^{n-1}&=&\ve\tau^2
\bigg\{\nu (x_1^n-x_2^n)-\alpha (x_2^n)^3\bigg\} ,\non
\eeqa
where $\theta$ is defined in Eq. \fr{nl-model-x-o}.
Setting $x_j^n\approx A_j^n\exp (-i\theta n) +\cc$ and following
the same procedure as in the case of the three- or more- site model, we
reach a regularized RG map up to ${\cal O}(\ve)$
\beq
(1-iT L_j )A_j^{n+1}=(1+iTL_j)\exp (iQ|A_j^{n}|) A_j^n, 
\quad (j=1,2)\lb{2disc_TQ}
\eeq
where an operator $L_j$ is defined by
\beq
L_1A_1^n\equiv -A_1^n+A_2^n,\qquad
L_2A_2^n\equiv A_1^n-A_2^n.
\lb{def-L_j}
\eeq
The detail derivation of this regularized RG map is shown in Appendix B.
It is worth noting that the map  \fr{2disc_TQ} has not only
symplectic symmetry but also a conserved quantity
\beq
\sum_{j=1,2}|A_j^n|^2=\sum_{j=1,2}|A_j^0|^2. \lb{conserv}
\eeq
The map  \fr{2disc_TQ} has   oscillating solutions
\beqa
A_1^n&=&A_2=A^0\exp\bigg(iQ|A^0|^2 n\bigg), \lb{tori-21}\\
&\mbox{or}& \non \\
A_1^n&=&-A_2=A^0\exp\bigg(iQ|A^0|^2 n\bigg),\lb{tori-22}
\eeqa
where $A^0$ is a constant.
The family of these solutions lies on a one-dimensional torus in the
phase space that spanned by $A_j^n$ and $A_j^{n~*}$.
The (modulational) stability of  solutions \fr{tori-21} and \fr{tori-22} is
of interest because the stability determines  hyperbolicity
or ellipticity of the torus.
If the torus is hyperbolic,  unstable and stable manifolds attach to the
torus like  ``whiskers'' \cite{Arn64} and  we call such unstable
and stable manifolds whiskers in this paper.

\begin{figure}[H]
\begin{center}
\psfrag{(a)}[][]{\tiny (a)}
\psfrag{(b)}[][]{\tiny (b)}
\psfrag{(c)}[][]{\tiny (c)}
\psfrag{i}[][]{\tiny iteration number $n$}
\psfrag{x_1^n}[][]{$x_1^n$}
\includegraphics[width=4.2cm]{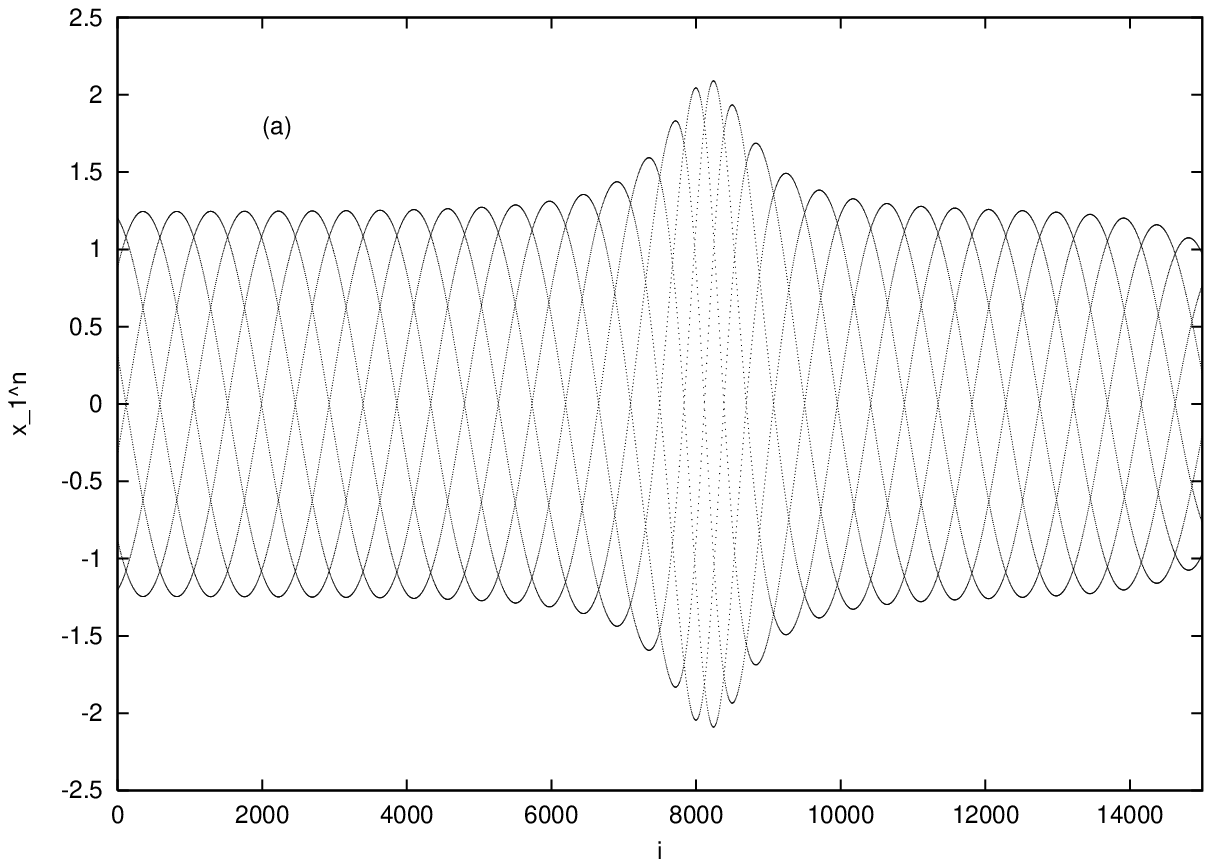}
\includegraphics[width=4.2cm]{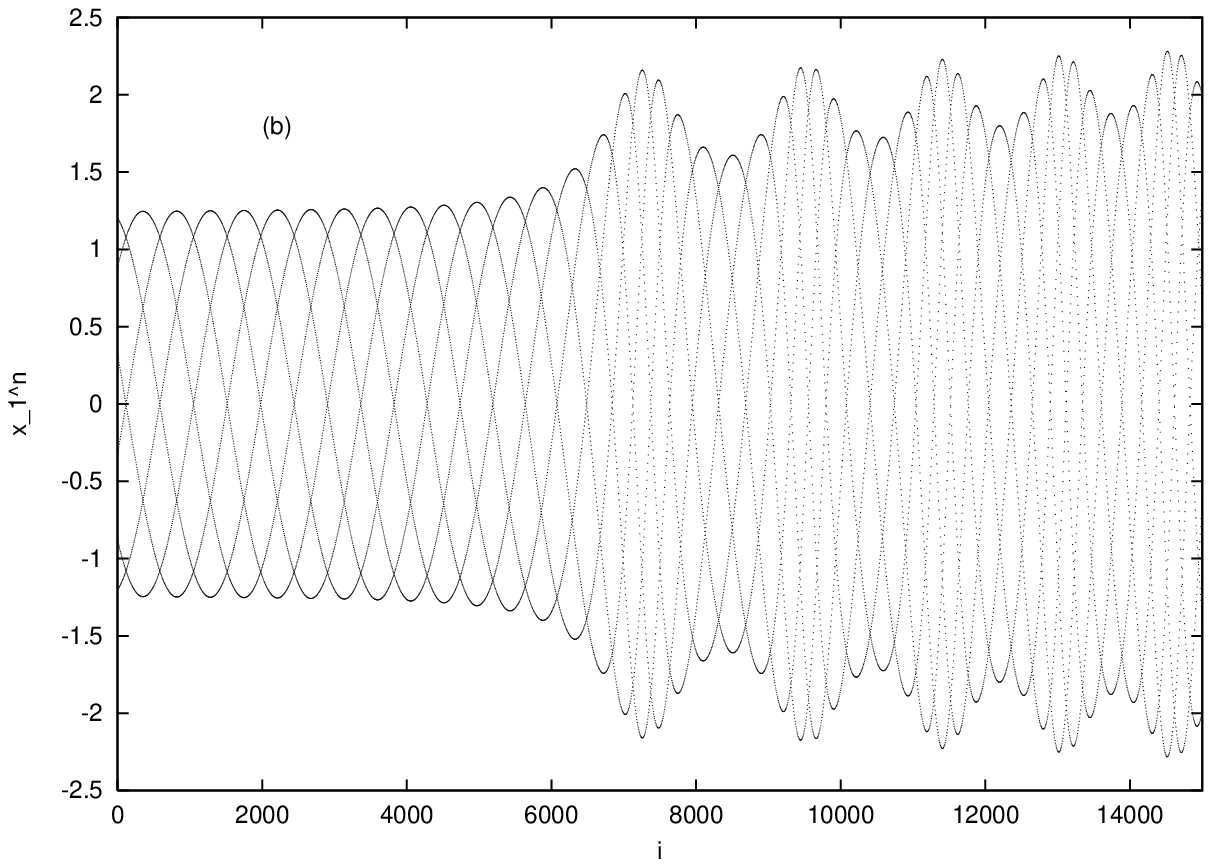}
\includegraphics[width=4.2cm]{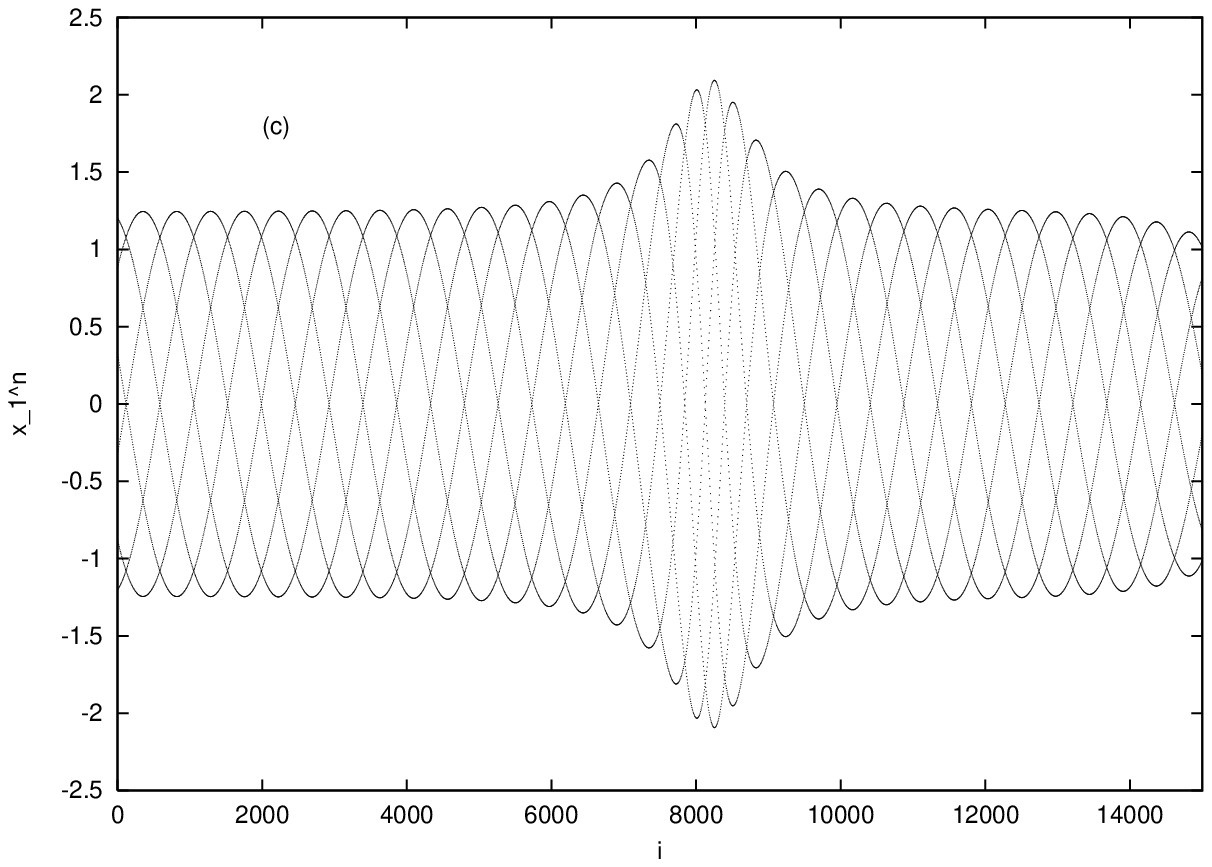}
\caption{
The time sequences of three-site model.  
(a) trajectory constructed from the original map, (b) the naive RG map 
and (c) the regularized RG map.
\label{fig:3comp_ex_nr_rr}
}
\end{center}
\end{figure}

Behavior of such  whiskers is the main object to investigate
in the following sections 3 and 4.
Let us show  effectiveness  of the regularized RG map by numerical
calculations.
In Fig. \ref{fig:3comp_ex_nr_rr}
  , trajectories obtained
from the naive RG map \fr{nl-model-nRGM}  and the regularized RG map
   \fr{nl-model-RRGM}  are depicted to be compared to an
exact trajectory of the original map
(Eqs. \fr{nl-model-x}--\fr{nl-model-p}).
All initial points are set very close to a hyperbolic torus.
The trajectory given by the regularized RG map is in good agreement with
the exact solution.

Owing to  the conserved quantity \fr{conserv}, the time-continuous
limit ($\tau \to 0$) of the RG map \fr{2disc_TQ} becomes integrable
and  analytical expression of whiskers, which is exactly homoclinic in this
case, is given in the Appendix C.

\section{Whiskered Tori }
In this section, we analyze the modulational instability of spatially uniform
oscillating solutions (tori) of the reduced maps derived in the previous
section
and introduce whiskered tori, that is,  the hyperbolic structure near
   periodic manifolds.  It is found that whiskers
retain the homoclinic structure approximately  near the onset of the
modulational instability,
where the homoclinic chaos is weak.

\subsection{Two-site case}
\begin{figure}[H]
\begin{center}
\psfrag{re}[][]{\tiny Re $a_j^n$}
\psfrag{im}[][]{\tiny Im $a_j^n$}
\psfrag{(a)}[][]{\tiny (a)}
\psfrag{(b)}[][]{\tiny (b)}
\psfrag{(0)}[][]{\tiny $(0)$}
\psfrag{(1)}[][]{\tiny $(1)$}
\psfrag{(2)}[][]{\tiny $(2)$}
\includegraphics[width=6cm]{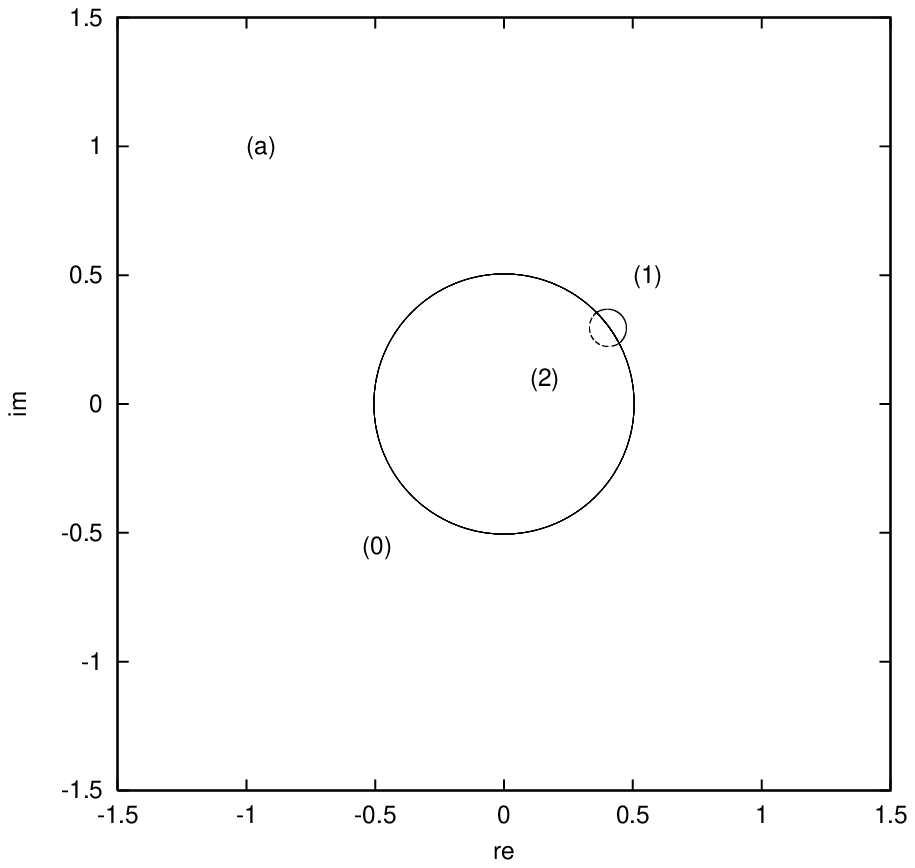}
\includegraphics[width=6cm]{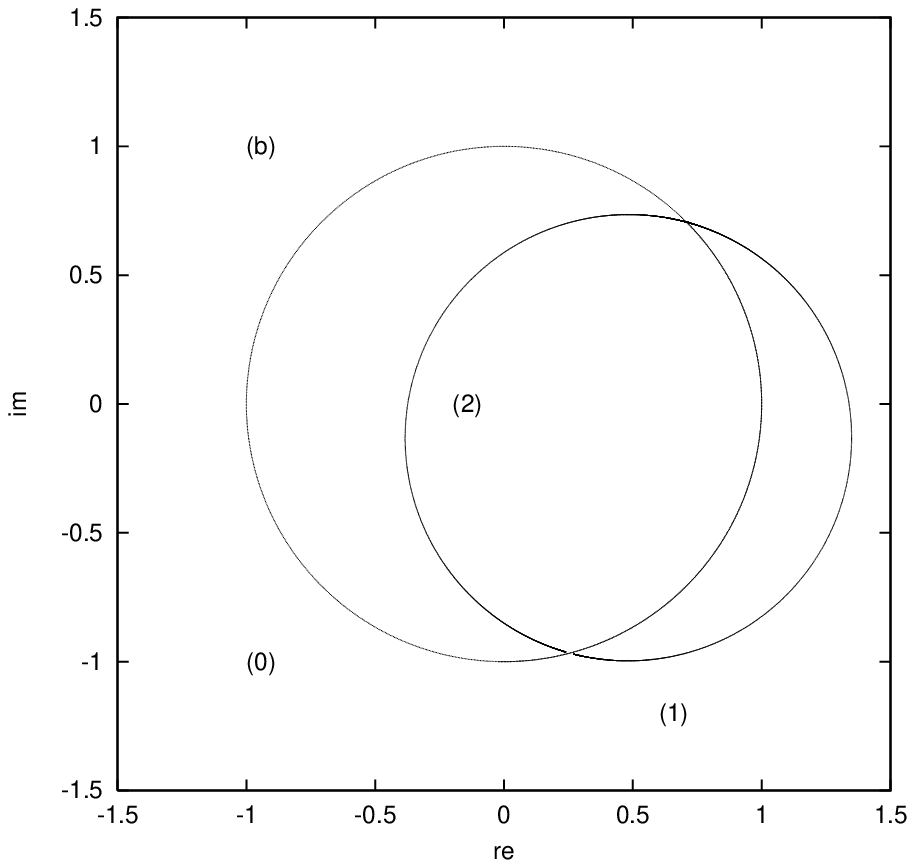}
\caption{
(a) The phase space of the two-site model near the hyperbolic torus for 
$|A^0|=|A_c|+0.005$.  
Here (0)  corresponds to a uniform solution $A_j^n=A^0\exp(iQ|A^0|^2n)$, 
and (1), (2) correspond to
$a_1^n$ and $a_2^n$, respectively. The initial condition is as follows : 
$\mbox{Re} A_1^0=|A^0|/\sqrt{2}+0.0002,
 \mbox{Re} A_2^0=|A^0|/\sqrt{2}+0.0001,
 \mbox{Im} A_1^0=|A^0|/\sqrt{2}+0.0,
 \mbox{Im} A_2^0=|A^0|/\sqrt{2}+0.0$.
(b) The initial conditions are same as (a) when  $|A^0|=|A_c|+0.5$.
\label{fig:2a}
}
\end{center}
\end{figure}

Let us analyze the reduced symplectic map with two sites,
whose continuous-time limit is integrable.\\
One of tori of Eq. \fr{2disc_TQ}
is just a spatially uniform  solution
$$A_j^n=A^0\exp(iQ|A^0|^2n),~(j=1,2).$$
To study the phase space structure around the solution,
we consider a perturbed solution of the form
\beq
A_j^n=A^0\exp(iQ|A^0|^2n)\bigg(1+\mu_j^n\bigg),\lb{2disc_perturb}
\eeq
where $|\mu_j^n|^2\ll 1$. Substitution of this expression into
\fr{2disc_TQ} and retaining first order terms in $\mu_j^n$ , yield
\beqa
\left(
\begin{array}{rl}
1+iT& -iT \\
   -iT& 1+iT
\end{array}
\right)
\left(
\begin{array}{c}
\mu_1^{n+1} \\
\mu_2^{n+1}
\end{array}
\right)
&=&
(1+iQ|A^0|^2)
\left(
\begin{array}{rl}
1-iT& iT \\
   iT& 1-iT
\end{array}
\right)
\left(
\begin{array}{c}
\mu_1^n \\
\mu_2^n
\end{array}
\right)\non\\
&&
+iQ|A^0|^2
\left(
\begin{array}{rl}
1-iT& iT \\
   iT& 1-iT
\end{array}
\right)
\left(
\begin{array}{c}
\mu_1^{n~*} \\
\mu_2^{n~*}
\end{array}
\right),\lb{2disc_linearli}
\eeqa
and  complex conjugate  to the above expression.
The eigenvalues of Eqs.\fr{2disc_linearli}  are
\beq
1,\quad 1,\quad \lambda_{\pm}=\beta\pm\sqrt{\beta^2-1} , \lb{eigen2}
\eeq
where $\beta\equiv (1-4T^2+4Q|A^0|^2T)/(1+4T^2)$.
We obtain conditions for the instability
\beq
|A^0|^2>\frac{2T}{Q},\lb{2disc_insta_1}
\eeq
or
\beq
|A^0|^2<-\frac{1}{2QT},
\eeq
which comes  from  $\beta^2>1$.
Hereafter, we concentrate on the condition \fr{2disc_insta_1}, under  which
the torus becomes hyperbolic for the amplitude $|A^0|$ larger than
a critical value ($\sqrt{2T/Q}$).
  From the spectrum of  eigenvalues  \fr{eigen2}, we find
that   whiskers consists of one-dimensional unstable manifold
(departing whisker)
and
one-dimensional stable manifold (arriving whisker).
To construct  whiskers numerically, 
we transform $A_j^n$ to $a_j^n  (j=1,2)$ where
$$
A_j^n=a_j^n\exp(iQ|A^0|^2 n).
$$
The torus with  departing whiskers are depicted in Figs \ref{fig:2a}.
Even if $|A^0|$ is considerably larger than the critical value,
   whiskers are found to be almost homoclinic and their behavior appears
to be regular.
This apparent regularity  reflects the integrability of continuous-time
limit of the map with two sites and is sharp contrast to the case of three or
more sites as studied in the subsequent sections.
Since difference between the map \fr{2disc_TQ} and its  integrable
continuous-time limit is very small (beyond all order), the weak
homoclinic chaos may be almost invisible in the present range of
numerical calculations.
\subsection{Three-site case}
Let us consider the regularized RG map
   with three sites \fr{nl-model-RRGM}.
   When the site number is odd,  general stability analysis is possible
and is provided in Appendix D.
Similar linearized analysis around a uniformly oscillation solution
as in the previous subsection
leads to
instability conditions:
\beq
|A^0|^2>\frac{3T}{Q},\lb{3site-insta1}
\eeq
or
\beq
|A^0|^2<\frac{-1}{3QT}.\lb{3site-insta2}
\eeq
As well as the two-site model, we consider the condition \fr{3site-insta1}
only. The spectrum of the linearized map is given as
\beqa
\lambda_a&\equiv&\beta_a\pm\sqrt{\beta_a^2-1},\qquad (a=0,1)\lb{spctrum3}\\
\beta_0&=&1,\qquad (\mbox{singlet})\non\\
\beta_1&=&\frac{1+6Q|A^0|^2T-9T^2}{1+9T^2},\qquad(\mbox{doublet}),\non
\eeqa

\begin{figure}[H]
\begin{center}
\psfrag{re}[][]{\tiny Re $a_j^n$}
\psfrag{im}[][]{\tiny Im $a_j^n$}
\psfrag{(a)}[][]{\tiny (a)}
\psfrag{(b)}[][]{\tiny (b)}
\psfrag{(0)}[][]{\tiny $(0)$}
\psfrag{(1)}[][]{\tiny $(1)$}
\psfrag{(2)}[][]{\tiny $(2)$}
\psfrag{(3)}[][]{\tiny $(3)$}
\includegraphics[width=6cm]{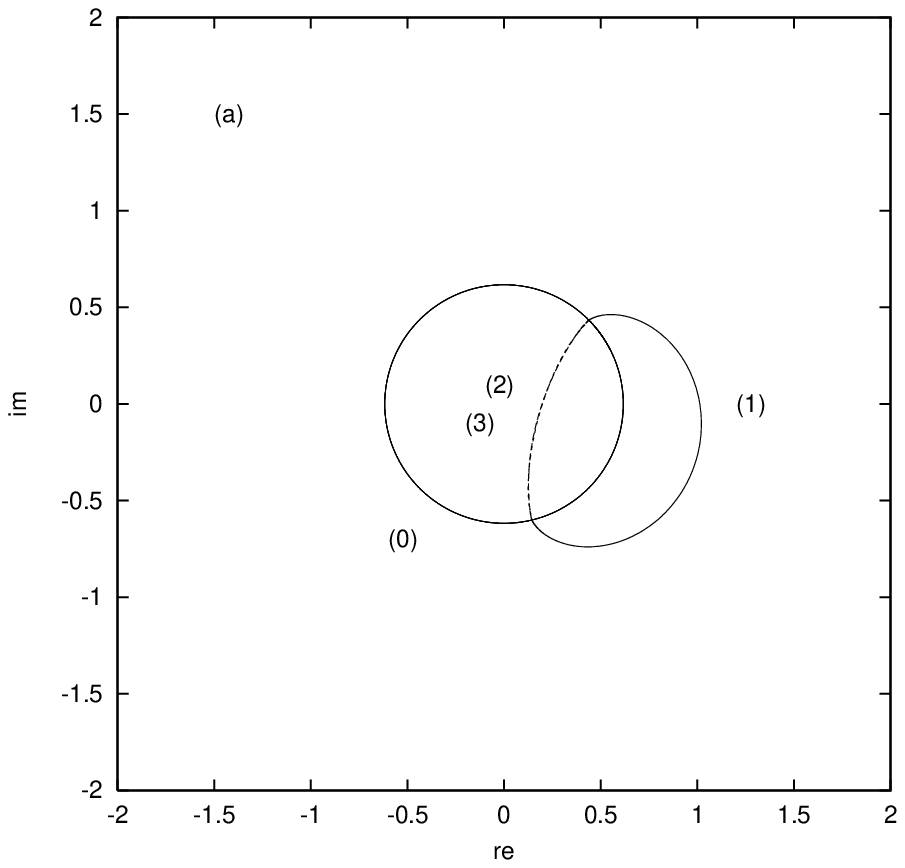}
\includegraphics[width=6cm]{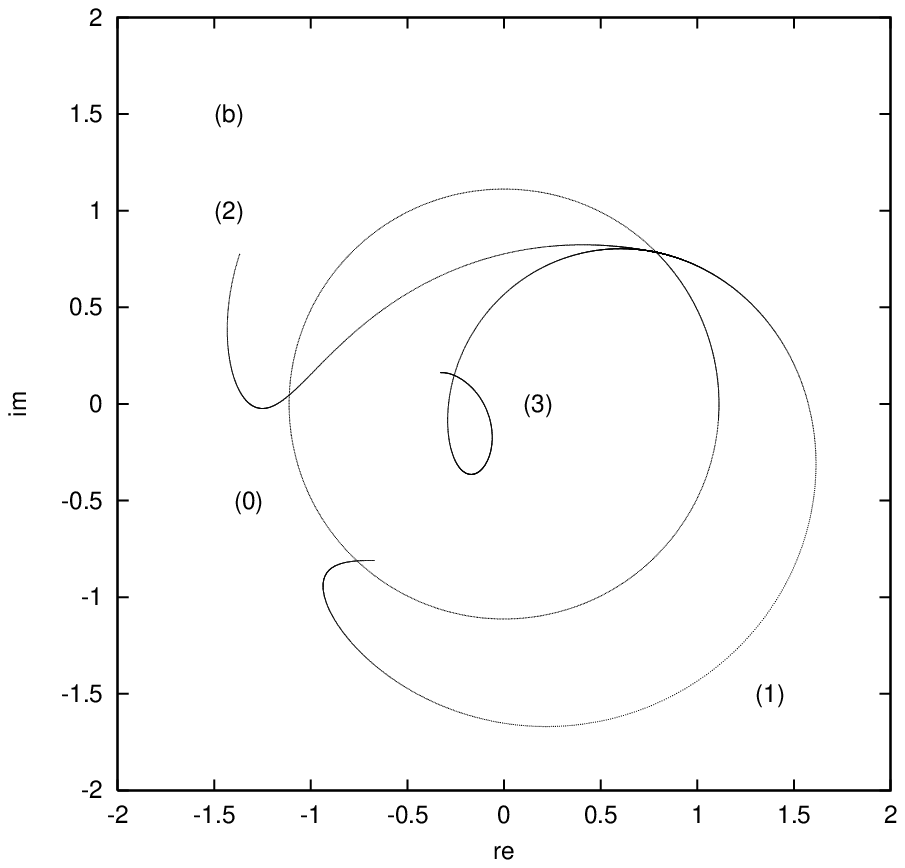}
\caption{
(a) The phase space of the three-site model near the hyperbolic torus for 
$|A^0|=|A_c|+0.005$.  
Here (0)  corresponds to a uniform solution $A_j^n=A^0\exp(iQ|A^0|^2n)$, 
and (1), $\cdots$, (3) correspond to
$a_1^n, \cdots, a_3^n$, respectively. The initial condition is as follows : 
$\mbox{Re} A_1^0=|A^0|/\sqrt{2}+0.0003,
 \mbox{Re} A_2^0=|A^0|/\sqrt{2}+0.0002,
 \mbox{Re} A_3^0=|A^0|/\sqrt{2}+0.0001,
 \mbox{Im} A_1^0=|A^0|/\sqrt{2}+0.0,
 \mbox{Im} A_2^0=|A^0|/\sqrt{2}+0.0,
 \mbox{Im} A_3^0=|A^0|/\sqrt{2}+0.0 $.
(b) The initial conditions are same as (a) when  $|A^0|=|A_c|+0.5$.
\label{fig:3a}
}
\end{center}
\end{figure}

The spectrum \fr{spctrum3} shows that
   the dimension of both departing and arriving whiskers are two for three-site
   case.
The tori with its unstable manifolds are depicted
in Figs. \ref{fig:3a}.
These figures show that the whiskers' behavior
is different from that of two-site case. The difference may be attributed to
one of integrability property of their continuous-time limit.
The greater is the amplitude of tori, the more complicate is
the whisker's behavior.  When the amplitude is just above a critical
value ($\sqrt{3T/Q}$) for the instability, whiskers are found to be
still homoclinic-like as shown in Fig \ref{fig:3a} (a).

\subsection{Four-site case}
   We analyse the modulational instability of a uniformly
   oscillating solution of the regularized RG map with four sites
   \fr{nl-model-RRGM} separately since general analysis is
   possible only for the odd-site case (Appendix D).

The linearized map around a
uniformly oscillating solution (a torus) has the following eight eigenvalues
\beqa
\lambda_{a\pm}&\equiv&\beta_a\pm\sqrt{\beta_a^2-1},\quad (a=0,1,2)\non\\
\beta_0 &=& 1 ,\qquad (\mbox{singlet})  \non\\
\beta_1 &=& \frac{1+4Q|A^0|^2 T-4T^2}{1+4T^2},\qquad (\mbox{doublet})\non\\
\beta_2 &=& \frac{1+8Q|A^0|^2 T-16T^2}{1+16T^2},\qquad (\mbox{singlet}).\non
\eeqa

The eigenvalues give the following instability conditions of the torus.
\beq
|A^0|^2>\frac{2T}{Q}\quad \mbox{(doublet)} ,
\qquad  |A^0|^2>\frac{4T}{Q}\quad \mbox{(singlet)},
\lb{4sites-condi-1}
\eeq
or
\beq
|A^0|^2<-\frac{1}{2QT}\quad \mbox{(doublet)},
\qquad |A^0|^2<-\frac{1}{4QT}\quad \mbox{(singlet)},
\lb{4sites-condi-2}
\eeq
where we consider the conditions  \fr{4sites-condi-1} only as before.
The spectrum of eigenvalues gives the dimension of  whiskers associated
with the torus:
\beqa
\bullet &&|A^0|^2<\frac{2T}{Q}\qquad \cdots\quad\mbox{no whiskers (elliptic)}
,\non\\
\bullet &&\frac{2T}{Q}<|A^0|^2<\frac{4T}{Q}
\cdots\quad\mbox{2 dimensional departing and 2 dim arriving }\non\\
&&\hspace{4cm}\mbox{whiskers and ellipticity is equal to 2 dimensions,}\non\\
\bullet &&|A^0|^2>\frac{4T}{Q}\qquad
\cdots\quad\mbox{3 dimensional departing and 3 dimensional}\non\\
&&\hspace{4cm}\mbox{arriving whiskers}.\non
\eeqa

\begin{figure}[H]
\begin{center}
\psfrag{re}[][]{\tiny Re $a_j^n$}
\psfrag{im}[][]{\tiny Im $a_j^n$}
\psfrag{(a)}[][]{\tiny (a)}
\psfrag{(b)}[][]{\tiny (b)}
\psfrag{(0)}[][]{\tiny $(0)$}
\psfrag{(1)}[][]{\tiny $(1)$}
\psfrag{(2)}[][]{\tiny $(2)$}
\psfrag{(3)}[][]{\tiny $(3)$}
\psfrag{(4)}[][]{\tiny $(4)$}
\includegraphics[width=6cm]{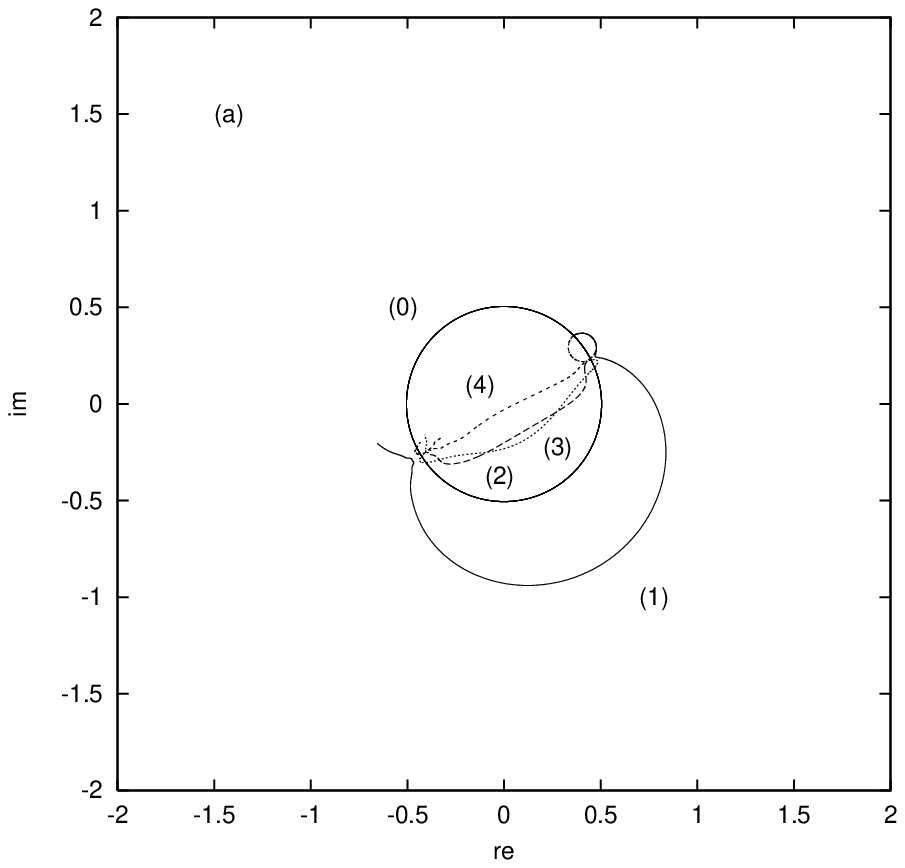}
\includegraphics[width=6cm]{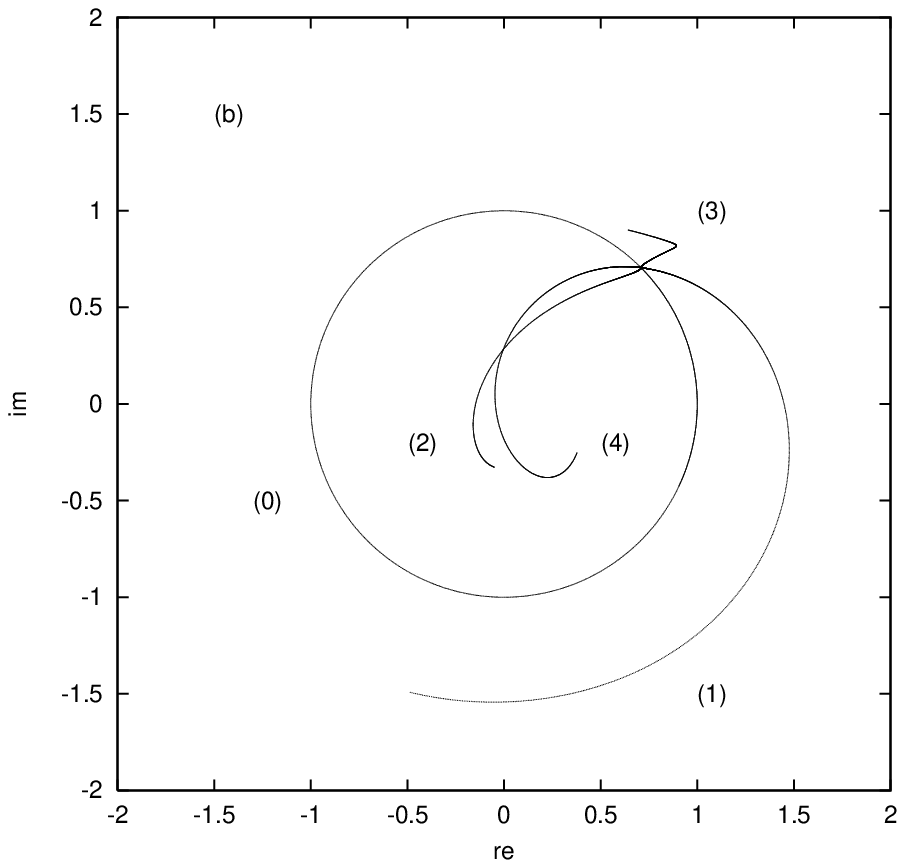}
\caption{
(a) The phase space of the four-site model near the hyperbolic torus for 
$|A^0|=|A_c|+0.005$.  
Here (0)  corresponds to a uniform solution $A_j^n=A^0\exp(iQ|A^0|^2n)$, 
and (1), $\cdots$, (4) correspond to
$a_1^n,\cdots, a_4^n$, respectively. The initial condition is as follows : 
$\mbox{Re} A_1^0=|A^0|/\sqrt{2}+0.0004,
 \mbox{Re} A_2^0=|A^0|/\sqrt{2}+0.0003,
 \mbox{Re} A_3^0=|A^0|/\sqrt{2}+0.0002,
 \mbox{Re} A_4^0=|A^0|/\sqrt{2}+0.0001,
 \mbox{Im} A_1^0=|A^0|/\sqrt{2}+0.0,
 \mbox{Im} A_2^0=|A^0|/\sqrt{2}+0.0,
 \mbox{Im} A_3^0=|A^0|/\sqrt{2}+0.0,
 \mbox{Im} A_4^0=|A^0|/\sqrt{2}+0.0,
 $.
(b) The initial conditions are same as (a) when  $|A^0|=|A_c|+0.5$.
\label{fig:4a}
}
\end{center}
\end{figure}

The torus with its unstable manifold is depicted in Fig. \ref{fig:4a}.
As similar to the three-site case,
the greater is the amplitude of torus, the more complicate is
whiskers' behavior.

\section{Random Rotation of Homoclinic-like Orbits }
\begin{figure}[H]
\begin{center}
\psfrag{re}[][]{\tiny Re $a_j^n$}
\psfrag{im}[][]{\tiny Im $a_j^n$}
\psfrag{(0)}[][]{\tiny $(0)$}
\psfrag{(1)}[][]{\tiny $(1)$}
\psfrag{(2)}[][]{\tiny $(2)$}
\psfrag{(3)}[][]{\tiny $(3)$}
\includegraphics[width=6cm]{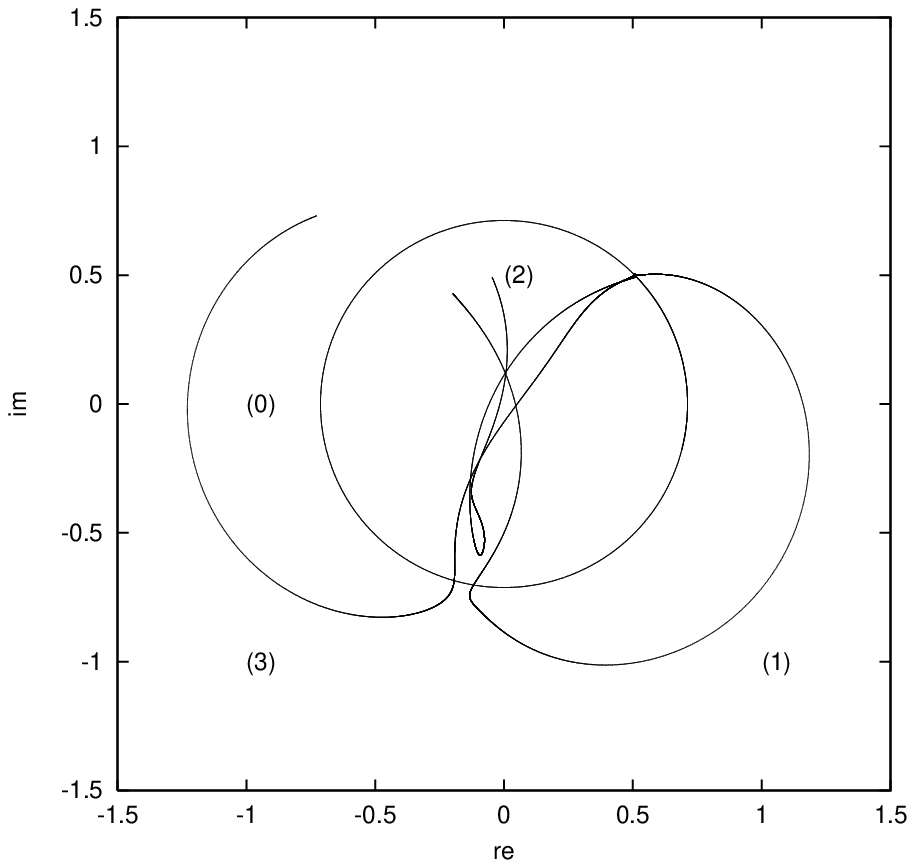}
\caption{
The phase space of the three-site model near the hyperbolic torus for 
$|A^0|=|A_c|+0.1$.  
Here (0)  corresponds to a uniform solution $A_j^n=A^0\exp(iQ|A^0|^2n)$, 
and (1), $\cdots$, (3) correspond to
$a_1^n,\cdots, a_3^n$, respectively. The initial condition is as follows : 
$\mbox{Re} A_1^0=|A^0|/\sqrt{2}+0.003,
 \mbox{Re} A_2^0=|A^0|/\sqrt{2}+0.002,
 \mbox{Re} A_3^0=|A^0|/\sqrt{2}+0.001,
 \mbox{Im} A_1^0=|A^0|/\sqrt{2}+0.0,
 \mbox{Im} A_2^0=|A^0|/\sqrt{2}+0.0,
 \mbox{Im} A_3^0=|A^0|/\sqrt{2}+0.0,
 $.
\label{fig:3a-homo}
}
\end{center}
\end{figure}

\begin{figure}[H]
\begin{center}
\psfrag{(a)}[][]{}
\psfrag{i}[][]{\tiny iteration number $n$}
\psfrag{(0)}[][]{\tiny $(0)$}
\psfrag{(1)}[][]{\tiny $(1)$}
\psfrag{(2)}[][]{\tiny $(2)$}
\includegraphics[width=8.2cm]{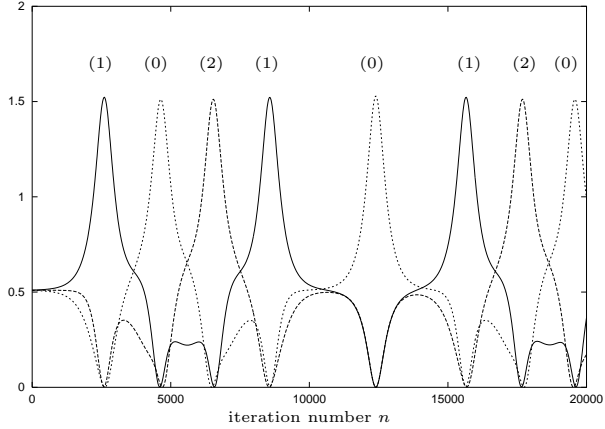}
\caption{
The time sequences of $|a_j^n|^2$ of the three-site model near the 
hyperbolic torus for 
$|A^0|=|A_c|+0.1$.  
Here (1), (2) and (0) correspond to $|a_1^n|^2$, $|a_2^n|^2$ and $|a_3^n|^2$,  
respectively. The orbit is characterized by the sequece of site-numbers 
$10210120 \cdots$.
The initial condition is as follows : 
$\mbox{Re} A_1^0=|A^0|/\sqrt{2}+0.003,
 \mbox{Re} A_2^0=|A^0|/\sqrt{2}+0.002,
 \mbox{Re} A_3^0=|A^0|/\sqrt{2}+0.001,
 \mbox{Im} A_1^0=|A^0|/\sqrt{2}+0.0,
 \mbox{Im} A_2^0=|A^0|/\sqrt{2}+0.0,
 \mbox{Im} A_3^0=|A^0|/\sqrt{2}+0.0. $
\label{fig:3sa01long}
}
\end{center}
\end{figure}

\begin{figure}[H]
\begin{center}
\psfrag{(b)}[][]{}
\psfrag{i}[][]{\tiny iteration number $n$}
\psfrag{(0)}[][]{\tiny $(0)$}
\psfrag{(1)}[][]{\tiny $(1)$}
\psfrag{(2)}[][]{\tiny $(2)$}
\psfrag{(3)}[][]{\tiny $(3)$}
\includegraphics[width=8.2cm]{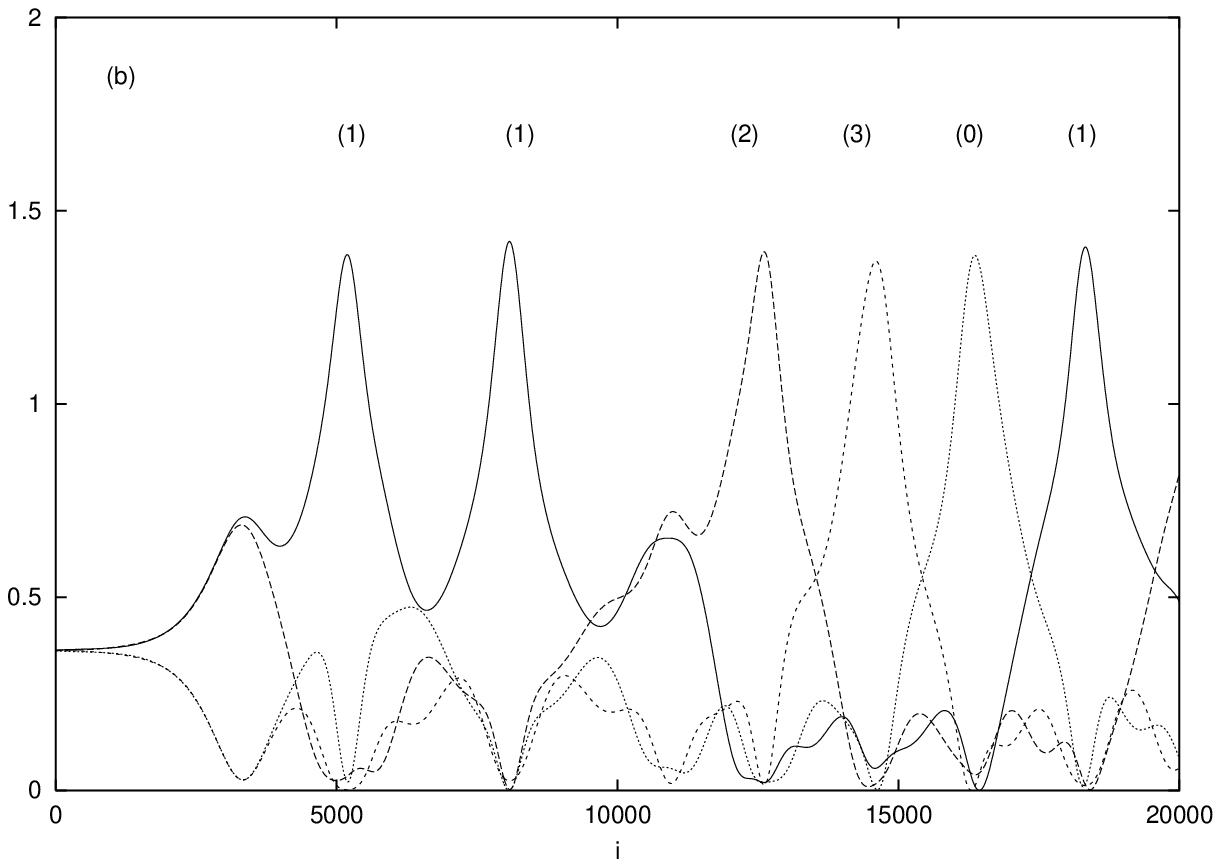}
\caption{
The time sequences of $|a_j^n|^2$ of the four-site model near the 
hyperbolic torus for 
$|A^0|=|A_c|+0.1$.  
Here (1), (2), (3) and (0) correspond to $|a_1^n|^2$, $|a_2^n|^2$, $|a_3^n|^2$ 
and $|a_4^n|^2$, respectively.
The orbit is characterized by the sequece of site-numbers
$ 112301 \cdots$.
The initial condition is as follows : 
$\mbox{Re} A_1^0=|A^0|/\sqrt{2}+0.004,
 \mbox{Re} A_2^0=|A^0|/\sqrt{2}+0.003,
 \mbox{Re} A_3^0=|A^0|/\sqrt{2}+0.002,
 \mbox{Re} A_3^0=|A^0|/\sqrt{2}+0.001,
 \mbox{Im} A_1^0=|A^0|/\sqrt{2}+0.0,
 \mbox{Im} A_2^0=|A^0|/\sqrt{2}+0.0,
 \mbox{Im} A_3^0=|A^0|/\sqrt{2}+0.0,
 \mbox{Im} A_4^0=|A^0|/\sqrt{2}+0.0, $
\label{fig:4sa01long}
}
\end{center}
\end{figure}

For slightly above the critical amplitude, we study  chaotic
orbit's behavior in the three- and four-site cases.
In such cases, an orbit wanders randomly  around
some homoclinic-like manifolds (whiskers) and it looks like
a random sequence of  homoclinic-like orbits
as shown in Fig. \ref{fig:3a-homo}.  
Each homoclinic-like orbit only differs in which site takes the
largest amplitude in the orbit.
In other word, a random rotation of site-numbers occurs in a sequence
of the same homoclinic-like structure.
This random rotation of site-numbers is understood by means of
the Bernoulli shift.
Let us characterize  each homoclinic-like orbit by the site-number of
the largest amplitude so that an orbit is represented by a sequence of
site-numbers
$$
10210120 \cdots \qquad (\mbox{\it resp}.\quad 112301 \cdots),\qquad \mbox{etc.}
$$
corresponding to the sequence in Figs. 
\ref{fig:3sa01long} and \ref{fig:4sa01long},
where $0$ indicates the site number $3$ or $4$.
This random sequence of numbers is conceived by the Bernoulli
sequence, which is generated by a simple map
$$
w^{n+1}=3w^{n}\qquad (\mbox{\it resp}.\quad w^{n+1}=4w^{n} ),\qquad \mbox{mod 1}.
$$
A periodic sequence of homoclinic-like orbits
corresponds to a periodic solution of the Bernoulli map
for a rational initial value on the interval $[0,1)$,
while a  random sequence is generated for an irrational initial value.

\begin{figure}[H]
\begin{center}
\psfrag{w^n}[][]{\tiny $w^n$}
\psfrag{w^{n+1}}[][]{\tiny $w^{n+1}$}
\psfrag{(a)}[][]{\tiny $(a)$}
\psfrag{(b)}[][]{\tiny $(b)$}
\includegraphics[width=6.2cm]{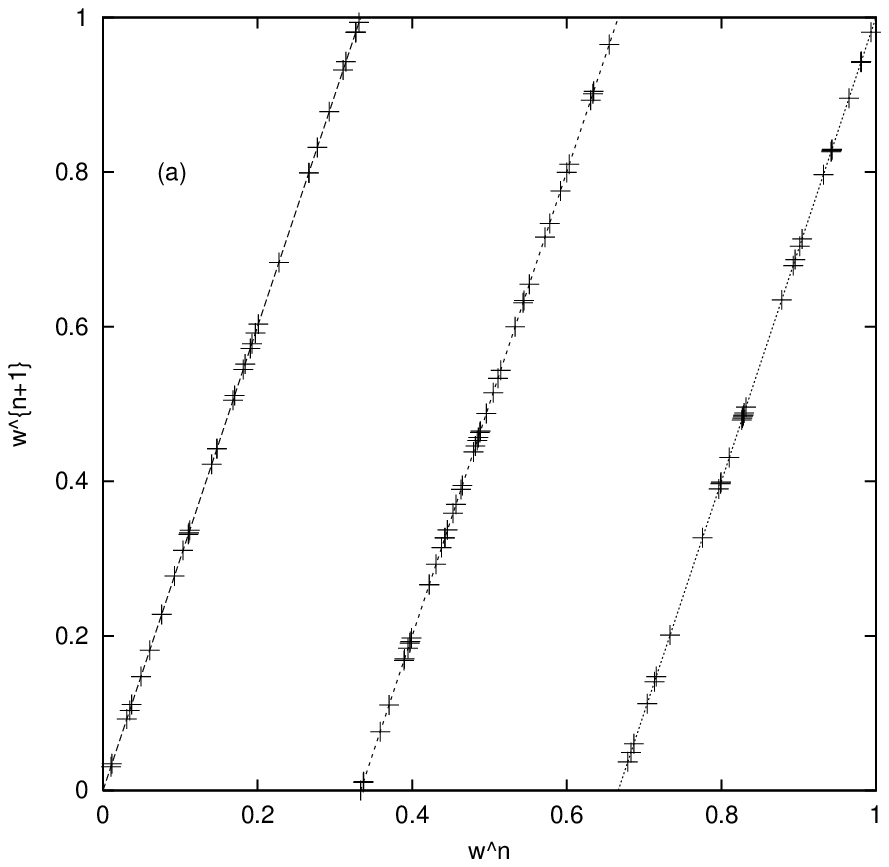}
\includegraphics[width=6.2cm]{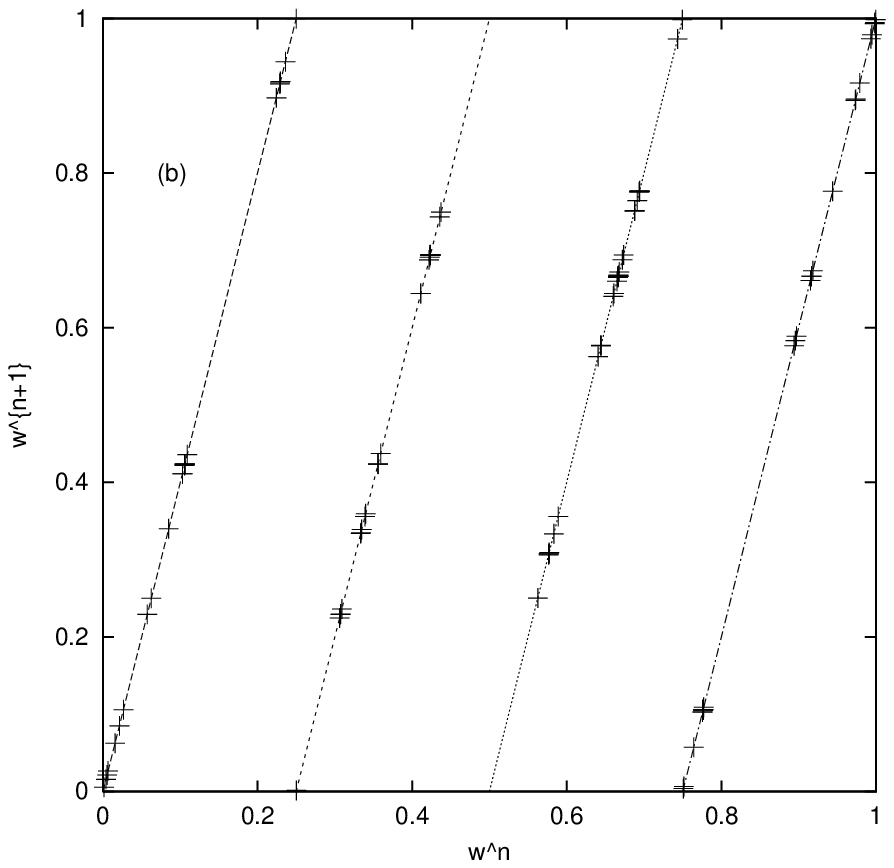}
\caption{
(a) The Bernouilli sequences of the three-site model near the 
hyperbolic torus for 
$|A^0|=|A_c|+0.1$.  
(b) same as (a) for  four-site model.
\label{fig:3_4sa01long_bs}
}
\end{center}
\end{figure}

Now, we present some number sequences of the homoclinic-like whiskers
obtained by numerical iterations of the three-site case and the four-site case
respectively  in Fig. \ref{fig:3_4sa01long_bs}.
These figures suggest that the sequences of homoclinic-like orbits
is not periodic but random.
It should be noted that such
a random rotation of  homoclinic-like orbits does not appear
   in the two-site case.

\section{Conclusion}
We present a regularization procedure to preserve the symplectic structure
of the RG map near a fully elliptic fixed point of a chain of weakly nonlinear
symplectic maps.
The regularization is accomplished by extended exponentiation of
the naive RG map and we derive a general reduced symplectic map
as an asymptotic RG map.

Analyzing the modulational instability of a uniformly oscillation solution of
   the reduced map, we find
a hyperbolic torus with  whiskers. We observe that  whiskers retain
a homoclinic-like structure in the case the amplitude of the uniformly
oscillating solution exceeds slightly the critical value.
For three or more-site cases, we find a random sequence of  the
homoclinic-like orbits.  The orbit is symbolized by a random rotation of
site-numbers generated by the Bernoulli shifts.
Although this chaos may be produced by
the homoclinic mechanism and is very weak,
it is easily visible as a random sequence of homoclinic-like orbits.
Such easily visible irregularity of homoclinic-like orbits may be general
in high-dimensional chaos and give an important visible tool to understand
   the onset of chaotic dynamics in high-dimensional spaces.

\section*{Acknowledgement}
The present work is, in part, supported by the Japan Society for Promotion
of Science, Grand-in-Aid for Scientific Research (C) 13640402.

%
\section*{Appendix}

\subsection*{Appendix A: Continuous-time  RG equation with three or more
sites}
Here, we give an RG equation of  continuous-time
   limit ($\tau\to 0$) of the weakly
nonlinear chain  \fr{nl-model-x} and \fr{nl-model-p},
which is  a spatially discretized nonlinear Schr\"odinger equation.
   Continuous-time limit of Eqs. \fr{nl-model-x} and
\fr{nl-model-p} reads
\beqa
\frac{dx_j}{dt}&=&p_j
=\frac{\partial H}{\partial p_j},\non\\
\frac{dp_j}{dt}&=&-\Omega^2x_j+\ve \bigg\{
\nu\bigg(x_{j+1}-2x_j+x_{j-1}\bigg)-\alpha x_j^3 \bigg\}
=- \frac{\partial H}{\partial x_j},\non\\
&& H=\sum_l \bigg[\frac{p_l^2+\Omega^2 x_l^2}{2}
+\ve\alpha\frac{x_l^4}{4}\bigg]+
\sum_l \ve\nu\bigg[\frac{(x_{l+1}-x_{l})^2}{2}\bigg].\non
\eeqa
Expanding $x_j$ as
$x_j=x_j^{(0)}+\ve x_j^{(1)}+{\cal O}(\ve^2 )$, a naive perturbed solution
is obtained as
\beqa
x_j^{(0)}&=&A_j\exp(-i\Omega t)+\cc\non\\
x_j^{(1)}&=&\frac{it}{2\Omega}
\bigg(\nu\sabun A_j-3\alpha |A_j|^2 A_j\bigg)\exp(-i\Omega t)
+\cc+\nr,\non
\eeqa
where $A_j$ is a complex valued integration constant for $j$-th site.
To remove secular terms  proportional to $t$, we define the renormalization
transformation $A_j\mapsto \tA_j(t)$ by  collecting all terms proportional
to the fundamental harmonic $\exp(-i\Omega t)$
$$
\tA_j(t)\equiv A_j+\ve\frac{it}{2\Omega}
\bigg(\nu\sabun A_j-3\alpha |A_j|^2 A_j\bigg)+{\cal O}(\ve^2),
$$
  From this transformation, the prescription of the RG method \cite{GMN99}
gives an RG equation
\beq
\frac{d\tA_j}{dt}=i\ve\frac{1}{2\Omega}
\Bigg(\nu\sabun \tA_j-3\alpha |\tA_j|^2 \tA_j\Bigg).
\lb{nonlinear-schro}
\eeq
   This RG equation automatically retains the symplectic property  and is known
   as  a spatially discretised nonlinear Schr\"odinger equation, which
is also non-integrable except the case of two sites.

\subsection*{Appendix B: Derivation of a reduced map of  two-site
nonlinear symplectic map}
In this subsection,  we derive a regularized RG map from
a nonlinear two-cite symplectic map chain.

Expanding $x_1^n$ and $x_2^n$ as $x_j^n=x_j^{n(0)}+\ve x_j^{n(1)}
+{\cal O}(\ve^2),(j=1,2)$, we have the following naive perturbative solution
of Eqs. \fr{2disc-orig-model-1} -- \fr{2disc-orig-model-3}
\beqa
x_j^{n(0)}&=&A_j\exp(-i\theta n)+\cc,\non\\
x_1^{n(1)}&=&\tau^2
\bigg\{
\nu (A_2-A_1)-3\alpha |A_1|^2A_1
\bigg\}\exp(-i\theta n)+\cc+\nr,\non\\
x_2^{n(1)}&=&\tau^2
\bigg\{
\nu (A_1-A_2)-3\alpha |A_2|^2A_2
\bigg\}\exp(-i\theta n)+\cc+\nr\non
\eeqa
The renormalization transformation $A_j\mapsto A_j^n$ to
remove the secular terms in the coefficient of $\exp(-i\theta n)$
is given by
\beqa
A_1^n\equiv A_1+\ve n\frac{i\tau^2}{2\sin\theta}
\bigg\{\nu(A_2-A_1)-3\alpha |A_1|^2 A_1 \bigg\}+{\cal O}(\ve^2) \non\\
A_2^n\equiv A_2+\ve n\frac{i\tau^2}{2\sin\theta}
\bigg\{\nu(A_1-A_2)-3\alpha |A_2|^2 A_2 \bigg\}+{\cal O}(\ve^2),
\lb{2disc-RGT}
\eeqa
from which we have a naive RG map up to ${\cal O}(\ve)$
\beq
A_j^n= A_j+\ve \frac{i\tau^2}{2\sin\theta}
\bigg(\nu L_j A_j-3\alpha |A_j^n|^2 A_j^n \bigg),\quad (j=1,2)
\lb{2disc-nRG}
\eeq
where, $L_j A_j$ is defined by Eq. \fr{def-L_j}.
The naive RG map \fr{2disc-nRG} should be regularized by means
the symplectic integration method.
Then,  we get a regularized RG map:
\beq
\bigg(1-iT L_j\bigg)A_j^{n+1}=
\bigg(1+iT L_j\bigg)
     \exp\bigg(iQ|A_j^n|^2\bigg)A_j ,
\lb{2disc-RRGM}
\eeq
which has the following  matrix form useful to numerical calculations:
\beqa
&&
\left(
\begin{array}{ccc}
1+iT &   -iT \\
   -iT & 1+iT  \\
\end{array}
\right)
\left(
\begin{array}{c}
A_1^{n+1} \\
A_2^{n+1} \\
\end{array}
\right)
=
\left(
\begin{array}{ccc}
1-iT &  +iT \\
   +iT & 1-iT \\
\end{array}
\right)
\left(
\begin{array}{c}
B_1^{n} \\
B_2^{n} \\
\end{array}
\right).\lb{matix-2}
\eeqa
By inverting the matrix in Eq.\fr{matix-2}, we have
$$
\left(
\begin{array}{c}
A_1^{n+1} \\
A_2^{n+1} \\
\end{array}
\right)
=\frac{1-2iT}{1+4T^2}
\left(
\begin{array}{ccc}
1   & 2iT  \\
2iT &1  \\
\end{array}
\right)
\left(
\begin{array}{c}
B_1^{n} \\
B_2^{n} \\
\end{array}
\right).
$$

\subsection*{ Appendix C:  Continuous-time RG equation with two sites}
In order to picture a homoclinic manifold resulting from the
modulational instability clearly,
let us analyze a continuous-time RG equation with two sites, which
is integrable.
In this case, the continuous-time RG equation has the following form
(see Eq.\fr{nonlinear-schro})
\beqa
\frac{d\tA_1}{dt}&=&\frac{i\ve}{2\Omega}
\bigg\{\nu(\tA_2-\tA_1)-3\alpha|\tA_1|^2\tA_1\bigg\},\non\\
\frac{d\tA_2}{dt}&=&\frac{i\ve}{2\Omega}
\bigg\{\nu(\tA_1-\tA_2)-3\alpha|\tA_2|^2\tA_2\bigg\},\non
\eeqa
Since  the regime where the modulational
instability occurs is of interest, we suppose $\nu=-|\nu|<0$ and $\alpha >0$
and introduce new variables:
$$
t'=|\nu|\frac{\ve}{2\Omega}t,\qquad Q_j
=\sqrt{\frac{3\alpha}{|\nu|}}\exp
\bigg\{\frac{i\nu\ve t}{2\Omega}\bigg\}\tA_j. \qquad (j=1,2)
$$
Then, we have
\beqa
&&\frac{dQ_1}{dt^{'}}=-i\bigg(Q_2+|Q_1|^2 Q_1\bigg)
=\frac{\partial {\cal H}}{\partial Q_1^* } ,\quad
\frac{d Q_1^* }{dt^{'}}=-\frac{\partial{\cal H}}{\partial Q_1},  \non\\
&&\frac{dQ_2}{dt^{'}}=-i\bigg(Q_1+|Q_2|^2 Q_2\bigg)
=\frac{\partial {\cal H}}{\partial Q_2^* } ,\quad
\frac{d Q_2^* }{dt^{'}}=-\frac{\partial {\cal H}}{\partial Q_2},\non\\
&&\qquad{\cal H}=-i\bigg\{ Q_1^*  Q_2+Q_1 Q_2^* +
(|Q_1|^4+|Q_2|^4)/2\bigg\},
\non
\eeqa
which has a uniformly oscillating  solution
$$
Q_j(t^{'})=Q_0(0)\exp\bigg\{-i(1+|Q_0(0)|^2)~t^{'}\bigg\} \quad (j=1,2).
$$
To construct  a homoclinic manifold of the unstable
periodic solution, we transform
$Q_j $ into $a_j $  as
$$
Q_j(t^{'})\equiv a_j(t^{'})\exp\bigg\{-i(1+|Q_0(0)|^2)~t^{'}\bigg\}.
$$
and have
\beqa
\frac{da_1}{dt^{'}}&=&-i\bigg[\bigg\{|a_1|^2-(1+|Q_0|^2)\bigg\}a_1+a_2
\bigg],\lb{2conti_a1}\\
\frac{da_2}{dt^{'}}&=&-i\bigg[\bigg\{|a_2|^2-(1+|Q_0|^2)\bigg\}a_2+a_1
\bigg].\lb{2conti_a2}
\eeqa
This system has two conserved quantities, i.e. the Hamiltonian and
$|a_1|^2+|a_2|^2$,  and is integrable.
After some manipulations of Eqs.\fr{2conti_a1} and \fr{2conti_a2},
we obtain an explicit expression of homoclinic manifold:
\beqa
a_1&=&b\sqrt{1+\Gamma}\exp\{i(\Theta +\Delta)/2\},\non\\
a_2&=&b\sqrt{1-\Gamma}\exp\{i(\Theta -\Delta)/2\},\non
\eeqa
and
\beqa
\Gamma(t')&=&\frac{2}{b^2}\sqrt{b^2-1}~\sech (-2\sqrt{b^2-1}t'),\non\\
\Delta(t')&=&\aCos\bigg(
\frac{1-b^2\Gamma^2/2}{\sqrt{1-\Gamma^2}}\bigg),\non\\
\Theta (t')&=&-\frac{1-b^2/2}{b^2-1}\frac{B}{\sqrt{1-B^2}}
\aTan\bigg\{\frac{B}{1-B^2}\tanh\bigg(2\sqrt{b^2-1}t'\bigg)\bigg\}
+\mbox{const}.\non ,
\eeqa
where $B\equiv 2\sqrt{b^2-1}/b^2$ and $b\equiv |Q_0(0)|>1$.

\subsection*{Appendix D: Modulational instability  for the case of
   odd  sites}
In this section, we provide  instability conditions for a uniformly
oscillating solution of the reduced map \fr{nl-model-RRGM} with odd
sites  by means of the Fourier analysis \cite{WH86}.
We substitute Eq. \fr{2disc_perturb} into Eq. \fr{nl-model-RRGM},
where  perturbation is set as
\beqa
\mu_j^n=\sum_a\hat{\mu}^j_a\exp\bigg(i\frac{2\pi}{N}aj\bigg),
\lb{mu-Fourier}
\eeqa
$$
a\in  \bigg\{-\frac{N-1}{2},-\frac{N-1}{2}+1,\cdots ,0,\cdots
,\frac{N-1}{2}-1,\frac{N-1}{2}\bigg\},
$$
and $N$ is an odd number ($N\ge 3$).
Assuming the periodic boundary condition
$\mu_{j+N}^n=\mu_{j}^n$, we have
a linearized equation for $\mu_j^n$.
\beq
\left(
\begin{array}{c}
\hat{\mu}_j^{n+1}\\
\hat{\mu}_j^{n+1~*}
\end{array}
\right)=
\left(
\begin{array}{rl}
d_a(1+iQ|A^0|^2)      & d_a iQ|A^0|^2\\
d_a^{-1} (-iQ|A^0|^2) & d_a^{-1}(1-iQ|A^0|^2)
\end{array}
\right)
\left(
\begin{array}{c}
\hat{\mu}_a^n\\
\hat{\mu}_a^{n~*}
\end{array}
\right),
\lb{linearl-Fourier}
\eeq
where $a\neq 0$ and
$$
d_a\equiv\frac{1-4iT\sin^2(\pi a/N)}{1+4iT\sin^2(\pi a/N)}.
$$
The eigenvalues of Eq. \fr{linearl-Fourier} are
\beq
\lambda_{a \pm}=\beta_a\pm\sqrt{\beta_a^2-1},
\lb{lambda_a-pm}
\eeq
where
$$
\beta_a\equiv
\frac{1+8QT|A^0|^2\sin^2(\pi a/N)-\bigg(4T\sin^2(\pi a/N)\bigg)^2}
{1+\bigg(4T\sin^2(\pi a/N)\bigg)^2}\in\bR.
$$
Note that $\lambda_{a +} \lambda_{a -}=1$.
The uniformly oscillating  solution is destabilized when
$$
\beta_a>1 \qquad  \mbox{or}\qquad\beta_a<1,
$$
that is
\beq
|A^0|^2>\frac{4T\sin^2(\pi a/N)}{Q},\qquad \mbox{or}\qquad
|A^0|^2<-\frac{1}{4QT\sin^2(\pi a /N)}.
\lb{insta-condition}
\eeq
The instability condition
   \fr{insta-condition} and the spectrum of
eigenvalues  \fr{lambda_a-pm} give 
information about dimensions of invariant manifolds around the uniformly
oscillating  solution (a torus) characterized by $|A^0|$.




\end{document}